\RequirePackage{lineno}
\documentclass[twocolumn,showpacs,aps,prd]{revtex4}

\usepackage{lineno}
\usepackage{graphicx}
\usepackage{dcolumn}
\usepackage{amsmath}
\usepackage{epsfig}
\usepackage{subfigure}
\usepackage{rotating}
\usepackage{enumerate}
\usepackage{comment}
\usepackage{hyperref}

\input pubboard/babarsym

\newcommand{\BABARPubYear}    {13}
\newcommand{\BABARPubNumber}  {015}

\newcommand{\SLACPubNumber} {15947}

\newcommand{\btosg}{\ensuremath{b\rightarrow s\gamma}}

\newcommand{\acp}{\ensuremath{A_{\CP}}\xspace}
\newcommand{\dacp}{\ensuremath{\Delta A_{X_s\gamma}}\xspace}
\newcommand{\xs}{\ensuremath{X_s}}
\newcommand{\BXsg}{\ensuremath{B \to X_s\gamma}}
\newcommand{\apeak}{\ensuremath{A_{\textrm{peak}}}\xspace}

\newcommand{\adet}{\ensuremath{A_{\textrm{det}}}\xspace}
\newcommand{\lambdase}{\ensuremath{\tilde{\Lambda}_{78}}\xspace}
\newcommand{\imes}{\ensuremath{\operatorname{Im}\left( C_{8g}/C_{7\gamma} \right)}\xspace}
\newcommand\TTT{\rule{0pt}{3.0ex}}       
\newcommand\BBB{\rule[-1.2ex]{0pt}{0pt}} 
\pacs{13.20.-v,13.25.Hw}

\def\figurebox#1#2#3{%
    \def\arg{#3}%
    \ifx\arg\empty
    {\hfill\vbox{\hsize#2\hrule\hbox to #2{\vrule\hfill\vbox to #1{\hsize#2\vfill}\vrule}\hrule}\hfill}%
    \else
    {\hfill\epsfbox{#3}\hfill}%
    \fi}

\begin{document}

\preprint{\babar-PUB-\BABARPubYear/\BABARPubNumber}
\preprint{SLAC-PUB-\SLACPubNumber}

\begin{flushleft}
\babar-PUB-\BABARPubYear/\BABARPubNumber\\
SLAC-PUB-\SLACPubNumber\\
\end{flushleft}

\title{
{
\large \bf
Measurements of Direct \CP Asymmetries in $\B \to  X_s \gamma$ decays \\ using Sum of Exclusive Decays}
}

%
\author{J.~P.~Lees}
\author{V.~Poireau}
\author{V.~Tisserand}
\affiliation{Laboratoire d'Annecy-le-Vieux de Physique des Particules (LAPP), Universit\'e de Savoie, CNRS/IN2P3,  F-74941 Annecy-Le-Vieux, France}
\author{E.~Grauges}
\affiliation{Universitat de Barcelona, Facultat de Fisica, Departament ECM, E-08028 Barcelona, Spain }
\author{A.~Palano$^{ab}$ }
\affiliation{INFN Sezione di Bari$^{a}$; Dipartimento di Fisica, Universit\`a di Bari$^{b}$, I-70126 Bari, Italy }
\author{G.~Eigen}
\author{B.~Stugu}
\affiliation{University of Bergen, Institute of Physics, N-5007 Bergen, Norway }
\author{D.~N.~Brown}
\author{L.~T.~Kerth}
\author{Yu.~G.~Kolomensky}
\author{M.~J.~Lee}
\author{G.~Lynch}
\affiliation{Lawrence Berkeley National Laboratory and University of California, Berkeley, California 94720, USA }
% ...
\collaboration{The \babar\ Collaboration}
\noaffiliation


\begin{abstract}

We measure the direct $C\!P$\xspace violation asymmetry, $A_{C\!P}\xspace$, in $B \xspace \rightarrow X_s\gamma$ and the isospin difference of the asymmetry, $\Delta A_{C\!P}$ \xspace, using 429 $\textrm{fb}^{-1}$ of data collected at $\Upsilon(4S)$ resonance with the \mbox{\slshape B\kern-0.1em{\smaller A}\kern-0.1em
    B\kern-0.1em{\smaller A\kern-0.2em R}}\: detector at the PEP-II asymmetric-energy $e^+e^-$ storage rings operating at the SLAC National Accelerator Laboratory. $B\xspace$ mesons are reconstructed from 10 charged $B\xspace$ final states and 6 neutral $B$ final states. We find $A_{C\!P}\xspace = +(1.7\pm1.9\pm1.0)\%$, which is in agreement with the Standard Model prediction and provides an improvement on the world average. Moreover, we report the first measurement of the difference
between $A_{C\!P}\xspace$ for charged and neutral decay modes, $\Delta A_{C\!P} \xspace = +(5.0\pm3.9\pm1.5)\%$. Using the value of $\Delta A_{C\!P}$ \xspace, we also provide 68\% and 90\% confidence intervals on the imaginary part of the ratio of the Wilson coefficients corresponding to the chromo-magnetic dipole and the electromagnetic dipole transitions.

\end{abstract}

\maketitle

\section{Introduction}
The flavor-changing neutral current decay $\BXsg$, where $\xs$ 
represents any hadronic system with one unit of strangeness, is highly suppressed in the standard model (SM), as is the direct \CP asymmetry, 
\begin{align}
	\label{eq:acpdef}
	\acp &= \frac{\Gamma_{\Bzb/\Bm \to X_{\s}\gamma}-\Gamma_{\Bz/\Bp \to X_{\sbar}\gamma}}{\Gamma_{\Bzb/\Bm \to X_{\s}\gamma}+\Gamma_{\Bz/\Bp \to X_{\sbar}\gamma}},
\end{align}
due to the combination of CKM and GIM suppressions~\cite{KGN}. New physics effects could enhance the asymmetry to a level as large as 15\%~\cite{HURTH}\cite{WOLF}\cite{CHUA}. 
The current world average of \acp\ based on the results from \babar~\cite{MILIANG}, Belle~\cite{BELLE} and CLEO~\cite{CLEO} is $-(0.8\pm2.9) \%$\cite{PDG}. The SM prediction for the asymmetry was found in a recent study to be long-distance-dominated~\cite{GIL} and to be in the range $-0.6\%<A_{CP}^{SM}<2.8\%$.

Benzke {\it et al.}~\cite{GIL} predict a difference in direct \CP asymmetry for charged and neutral \B mesons: 
\begin{align}
	\dacp &= A_{\Bpm\to X_s\gamma} - A_{\Bz/\Bzb \to X_s\gamma},
\end{align}
which suggests a new test of the SM. The difference, $\dacp$, arises from an interference term in \acp that depends on the charge of the spectator quark. The magnitude of $\dacp$ is proportional to \imes where $C_{7\gamma}$ and $C_{8g}$ are Wilson coefficients corresponding to the electromagnetic dipole  and the chromo-magnetic dipole transitions, respectively. The two coefficients are real in the SM; therefore $\dacp$=0. New physics contributions from the enhancement of the \CP-violating phase or of the magnitude of the two Wilson coefficients~\cite{KGN}\cite{Jung:2012vu}, or the introduction of new operators~\cite{Shimizu:2012zw} could enhance \dacp to be as large as 10\%~\cite{GIL}. Unlike $C_{7\gamma}$, $C_{8g}$ currently does not have a strong experimental constraint~\cite{Altmannshofer:2011gn}. Thus a measurement of \dacp together with the existing constraints on $C_{7\gamma}$ can provide a constraint on $C_{8g}$.

Experimental studies of $\BXsg$ are approached in one of two ways.
The inclusive approach relies entirely on observation of the high-energy
photon from these decays without reconstruction of the hadronic system
$\xs$.  By ignoring the $\xs$ system, this approach is sensitive to the full 
$\btosg$ decay rate and is robust against final state fragmentation effects.  
The semi-inclusive approach reconstructs the $\xs$ system in as many
specific final state configurations as practical.  This approach provides 
additional information, but since not all $\xs$ final states 
can be reconstructed without excessive background, fragmentation 
model-dependence is introduced if semi-inclusive measurements 
are extrapolated to the complete ensemble of $\BXsg$ decays.
$\babar$ has recently published results on the $\BXsg$ 
branching fraction and photon spectrum for both approaches~\cite{Lees:2012wg}\cite{Lees:2012ufa}. The inclusive approach has also been used to search for direct $\CP$ violation, but since the inclusive method does not distinguish
hadronic final states, decays due to $b \to d \gamma $ transitions
are included.

We report herein a measurement of $\acp$ and the first 
measurement of $\dacp$ using the semi-inclusive
approach with the full $\babar$ data set. We reconstruct 38 exclusive \B-decay modes, listed in Table~\ref{tab:xsmodes}, but 
for use in this analysis a subset of 16 modes (marked with an asterisk in Table~\ref{tab:xsmodes}) 
is chosen for which high statistical significance is achieved.  Also,
for this analysis, modes must be flavor self-tagging (i.e., the bottomness can be determined from the reconstructed final state).  
The 16 modes include ten charged $\B$ and six neutral $\B$ decays.  After all 
event selection criteria are applied, the mass of the hadronic $\xs$ system ($m_{\xs}$) in this measurement covers the range of about 0.6 to 2.0 \gevcc. 
The upper edge of this range approximately corresponds to 
a minimum photon energy in the $\B$ rest frame of 2.3 \gev. For $\BXsg$ decays with
$0.6 < m_{\xs} < 2.0 \gevcc$, the 10 charged \B modes used account for about
52\% of all $\Bp \to \xs \gamma$ decays and the six neutral modes 
account for about 34\% of all neutral $\Bz \to \xs \gamma$ decays~\cite{KLFraction}.  
In this analysis it is assumed that \acp and \dacp are independent
of final state fragmentation.  That is, it is assumed that \acp and
\dacp are independent of the specific $\xs$ final states used
for this analysis and independent of the $m_{\xs}$ distribution of the
selected events.

\begin{table}[ht]
\begin{center}
\caption{\label{tab:xsmodes} The 38 final states we reconstruct in this analysis. Charge conjugation is implied. The 16 final states used in the \CP measurement are marked with an asterisk.}
\begin{tabular}{l l |l  l}
\hline
\#	\BBB \TTT	& Final State							&\#		&Final State\\
\hline
\hline
1*		& $B^{+}\rightarrow K_{S}\pi^{+}\gamma$				& 20		& $B^{0}\rightarrow K_{S}\pi^{+}\pi^{-}\pi^{+}\pi^{-}\gamma$\\
2*		& $B^{+}\rightarrow K^{+}\pi^{0}\gamma$				& 21 		& $B^{0}\rightarrow K^{+}\pi^{+}\pi^{-}\pi^{-}\pi^{0}\gamma$\\
3*		& $B^{0}\rightarrow K^{+}\pi^{-}\gamma$				& 22 		& $B^{0}\rightarrow K_{S}\pi^{+}\pi^{-}\pi^{0}\pi^{0}\gamma$\\
4		& $B^{0}\rightarrow K_{S}\pi^{0}\gamma$				& 23*		& $B^{+}\rightarrow K^{+}\eta\gamma$\\
5*		& $B^{+}\rightarrow K^{+}\pi^{+}\pi^{-}\gamma$			& 24 		& $B^{0}\rightarrow K_{S}\eta\gamma$\\
6*		& $B^{+}\rightarrow K_{S}\pi^{+}\pi^{0}\gamma$			& 25 		& $B^{+}\rightarrow K_{S}\eta\pi^{+}\gamma$\\
7*		& $B^{+}\rightarrow K^{+}\pi^{0}\pi^{0}\gamma$			& 26		& $B^{+}\rightarrow K^{+}\eta\pi^{0}\gamma$\\
8		& $B^{0}\rightarrow K_{S}\pi^{+}\pi^{-}\gamma$			& 27*		& $B^{0}\rightarrow K^{+}\eta\pi^{-}\gamma$\\
9*		& $B^{0}\rightarrow K^{+}\pi^{-}\pi^{0}\gamma$			& 28		& $B^{0}\rightarrow K_{S}\eta\pi^{0}\gamma$\\
10		& $B^{0}\rightarrow K_{S}\pi^{0}\pi^{0}\gamma$			& 29		& $B^{+}\rightarrow K^{+}\eta\pi^{+}\pi^{-}\gamma$\\
11*		& $B^{+}\rightarrow K_{S}\pi^{+}\pi^{-}\pi^{+}\gamma$		& 30 		& $B^{+}\rightarrow K_{S}\eta\pi^{+}\pi^{0}\gamma$\\
12*		& $B^{+}\rightarrow K^{+}\pi^{+}\pi^{-}\pi^{0}\gamma$		& 31		& $B^{0}\rightarrow K_{S}\eta\pi^{+}\pi^{-}\gamma$\\
13*		& $B^{+}\rightarrow K_{S}\pi^{+}\pi^{0}\pi^{0}\gamma$		& 32		& $B^{0}\rightarrow K^{+}\eta\pi^{-}\pi^{0}\gamma$\\
14*		& $B^{0}\rightarrow K^{+}\pi^{+}\pi^{-}\pi^{-}\gamma$		& 33*		& $B^{+}\rightarrow K^{+}K^{-}K^{+}\gamma$\\
15		& $B^{0}\rightarrow K_{S}\pi^{0}\pi^{+}\pi^{-}\gamma$		& 34		& $B^{0}\rightarrow K^{+}K^{-}K_{S}\gamma$\\
16*		& $B^{0}\rightarrow K^{+}\pi^{-}\pi^{0}\pi^{0}\gamma$		& 35		& $B^{+}\rightarrow K^{+}K^{-}K_{S}\pi^{+}\gamma$\\
17		& $B^{+}\rightarrow K^{+}\pi^{+}\pi^{-}\pi^{+}\pi^{-}\gamma$	& 36		& $B^{+}\rightarrow K^{+}K^{-}K^{+}\pi^{0}\gamma$\\
18		& $B^{+}\rightarrow K_{S}\pi^{+}\pi^{-}\pi^{+}\pi^{0}\gamma$	& 37*		& $B^{0}\rightarrow K^{+}K^{-}K^{+}\pi^{-}\gamma$\\
19		& $B^{+}\rightarrow K^{+}\pi^{+}\pi^{-}\pi^{0}\pi^{0}\gamma$	& 38		& $B^{0}\rightarrow K^{+}K^{-}K_{S}\pi^{0}\gamma$\\
\hline
\end{tabular}
\end{center}
\end{table}

\section{Analysis Overview}

With data from the $\babar$ detector (Section~\ref{sec:detector}), we reconstructed \B candidates from various final states (Section~\ref{sec:breco}). We then trained two multivariate classifiers (Section~\ref{sec:selection}): one to separate correctly reconstructed \B decays from mis-reconstructed events and the other to reject the continuum background, $\ep e^- \to \qqbar$, where $q=u,d,s,c$. The output of the first classifier is used to select the best \B candidate for each event. Then, the outputs from both classifiers are used to reject backgrounds. We use the remaining events to determine the asymmetries.

We use identical procedures to extract three asymmetries: the asymmetries of charged and neutral B mesons, and of the combined sample, and the difference, $\dacp$. The bottomness of the $\B$ meson is
determined by the charge of the kaon for $\Bz$ and $\Bzb$, and by the total charge
of the reconstructed $\B$ meson for \Bp and $\Bm$.

We can decompose \acp into three components:
\begin{align}
	\acp = \apeak - \adet + D
	\label{eq:acpapeakrelation}
\end{align}
where \apeak is the fitted asymmetry of the events in the peak of the $\mes$ distribution (Section \ref{sec:apeak}), \adet is the detector asymmetry due to the difference in $K^+$ and $K^-$ efficiency (Section~\ref{sec:adet}), and $D$ is the bias due to peaking background contamination (Section~\ref{sec:dilution}). In this analysis we establish
upper bounds on the magnitude of $D$, and then treat
those as systematic errors.

\section{Detector and Data}
\label{sec:detector}
We use a data sample of 429 $\textrm{fb}^{-1}$~\cite{Lees:2013rw} collected at the $\Upsilon(4S)$ resonance, $\sqrt{s}=10.58\gevcc$, with the \babar\ detector at the PEP-II asymetric-energy $\B$ factory at the SLAC National Accelerator Laboratory. The data corresponds to $471 \times 10^6$ produced $\B\Bbar$ pairs. 

The \babar\: detector and its operation are described in detail elsewhere~\cite{Aubert:2001tu}\cite{BABARNEWNIM}. The charges and momenta of charged particles are measured by a five-layer double-sided silicon strip detector (SVT) and a 40-layer drift chamber (DCH) operated in a 1.5 T solenoidal field. Charged $K/\pi$ separation is achieved using $dE/dx$ information from the trackers and  by a detector of internally reflected Cherenkov light (DIRC), which measures the angle of the Cherenkov radiation cone. An electromagnetic calorimeter (EMC) consisting of an array of CsI(Tl) crystals measures the energy of photons and electrons.

We use a Monte Carlo (MC) simulation based on EvtGen~\cite{EVTGEN} to optimize the event selection criteria.  We model the background as $e^+e^-\to \qqbar$, $e^+ e^- \to \tau^+\tau^-$ and $\B\Bbar$.  We generate signal $\B\to X_s \gamma$ with a uniform photon spectrum and then weight signal MC events so that the photon spectrum matches the kinematic-scheme model~\cite{BBU} with parameter values consistent with the previous \babar\: $\B\to X_s \gamma$ photon spectrum analysis ($m_b =4.65 \gevcc$ and $\mu_\pi^2=0.20 \gev^2$)~\cite{babar_measurement}. We use JETSET~\cite{JETSET} as the fragmentation model and GEANT4~\cite{GEANT} to simulate the detector response.

\section{\B Reconstruction}
\label{sec:breco}
We reconstructed \B meson candidates from 38 final states listed in Table~ \ref{tab:xsmodes}. The 16 modes marked with an asterisk (*) in Table~ \ref{tab:xsmodes} are used in the \CP measurement. The other final states are either not flavor-specific final states or are low in yield. We reconstruct the unused modes  in order to veto them after selecting the best candidate. 
In total, we use 10 charged $\B$ final states and 6 neutral $\B$ final states in the \acp\ measurement. These final states are the same as those used in a previous $\babar$ analysis~\cite{MILIANG}.

Charged kaons and pions are selected from tracks classified with an error-correcting output code algorithm~\cite{BABARNEWNIM}\cite{ECOC}. The classification uses SVT, DIRC, DCH, and EMC information. The kaon particle identification (PID) algorithm has approximately 90\% efficiency and a pion-as-kaon misidentification rate of about 1\%. Pion identification is roughly 99\% efficient with a 15\% kaon-as-pion misidentification rate.

Neutral kaons are reconstructed from the decay $\KS \to \pi^+\pi^-$. The invariant mass of the two oppositely charged tracks is required to be between 489 and 507 MeV. The flight distance of the $\KS$ must be greater than 0.2~cm from the interaction point. The flight significance (defined as the  flight distance divided by the uncertainty in the flight distance) of the \KS must be greater than three. \KL and $\KS\to \pi^0\pi^0$ decays are not reconstructed for this analysis.

The neutral $\piz$ and $\eta$ mesons are reconstructed from two photons. We require each photon to have energy of at least 30 \mev for $\piz$ and at least 50 \mev for $\eta$. The invariant mass of the two photons must be in the range of [115,150] \mev for $\pi^0$ candidates and in the range of [470,620] \mev for $\eta$ candidates. Only $\pi^0$ and $\eta$ candidates with momentum greater than 200 \mev are used. We do not reconstruct $\eta \to \pi^+\pi^-\pi^0$ decays explicitly, but some are included in final states that contain $\pi^+\pi^-\pi^0$.

Each event is required to have at least one photon with energy $1.6 < E^*_\gamma < 3.0 \gev$, where the asterisk denotes variables measured in the  $\Upsilon(4S)$ center-of-mass (CM) frame. These photons are used as the primary photon in reconstructing $\B$ mesons. Such a photon must have a lateral moment~\cite{lateralmoment} less than 0.8 and the nearest EMC cluster must be at least 15 cm away. The angle of the photon momentum with respect to the beam axis must satisfy $-0.74<\cos \theta<0.93$.

The invariant mass of $X_s$(all daughters of the $\B$ candidate excluding the primary photon) must satisfy $0.6<m_{X_s}<3.2 \gevcc$. The $X_s$ candidate is then combined with the primary photon to form a $\B$ candidate, which is required to have an energy-substituted mass $\mes=\sqrt{s/4 - {p^*_B}^2}$, where $p^*_B$ is the momentum of $\B$ in the CM frame, greater than 5.24~\gevcc. We also require the difference between half of the beam total energy and the energy of the reconstructed $\B$ in the CM frame, $|\Delta E| = |E^*_\textrm{beam}/2 -E^*_B|$, to be less than 0.15~\gev. The angle between the thrust axis of the rest of the event(ROE) and the primary photon must satisfy $|\cos \theta^*_{T\gamma}| < 0.85$.

\section{Event and Candidate Selection}
\label{sec:selection}

There are three main sources of background.  The dominant source is continuum background, $\ep e^- \to \qqbar$. These events are more jet-like than the $e^+e^-\to \Upsilon(4S) \to \B\Bbar$. Thus, event shape variables provide discrimination. The continuum $\mes$ distribution does not peak at the $\B$ meson mass. The second background source is $\B\Bbar$ decays to final states other than $X_s\gamma$; hereafter we refer to these as generic $\B\Bbar$ decays. The third source is cross-feed background which comes from actual $\B \to X_s\gamma$ decays in which we fail to reconstruct the $\B$ in the correct final state. The $e^+e^- \to \tau^+\tau^-$ contribution is negligibly small.

We first place a preliminary selection on the ratio of angular moments~\cite{AngularMoment}\cite{Aubert:2002jb}, $L_{12}/L_{10} < 0.46$ to reduce the number of the continuum background events. This ratio measures the jettiness of the event. Since the mass of the $\B$ meson is close to half the mass of the $\Upsilon(4S)$, the kinetic energy that the $\B$ meson can have is less than that available to $e^+e^- \to$ light quark pairs. Therefore, the signal peaks at a lower value of $L_{12}/L_{10}$ than does the continuum background.

The $\B$ meson reconstruction typically yields multiple $\B$ candidates per event. To select the best candidate, we train a random forest classifier~\cite{RF} based on $\Delta E/\sigma_E$, where $\sigma_E$ is the uncertainty on the $\B$ candidate energy, the thrust of the reconstructed $\B$ candidate~\cite{Thrust}, $\pi^0$ momentum, the invariant mass of the $X_s$ system, and the zeroth and fifth Fox-Wolfram moments~\cite{FWmoments}.  This Signal Selecting Classifier (SSC) is trained on a large MC event sample to separate correctly reconstructed $\BXsg$ decays from mis-reconstructed ones. For each event, the candidate with the maximum classifier output is chosen as the best candidate. This is the main difference from a previous \babar\ analysis~\cite{MILIANG} which chose the event with the smallest $|\Delta E|$ as the best candidate. This method increases the efficiency by a factor 
of approximately two for the same misidentification rate. 

It should be emphasized that the best candidate selection procedure also selects final states in which the bottomness of the \B cannot be deduced from the final decay products (flavor-ambiguous final states). After selecting the best candidate, we keep only events in which the best candidate is reconstructed with the final states marked with an asterisk in Table~\ref{tab:xsmodes}. This removes events which are flavor-ambiguous final states from the \acp\ measurement. Furthermore, because of the way the SSC was trained to discriminate against mis-reconstructed $\B$ candidates, SSC also provides good discriminating power against the generic $\B\Bbar$ background.

To further reduce the continuum background we build another random forest classifier, the Background Rejecting Classifier (BRC), using the following variables: 
\begin{itemize}
\item $\pi^0$ score: the output from a random forest classifier using the invariant mass of the primary photon with all other photons in the event and the energy of the other photons, which is trained to reject high-energy photons that come from the $\pi^0\to\gamma\gamma$ decays.
\item Momentum flow~\cite{MomentumFlow} in $10^{\circ}$ increments about the reconstructed $\B$ direction.
\item Zeroth, first and second order angular moments along the primary photon axis computed in the CM frame of the ROE. 
\item The ratio of the second and the zeroth angular moments described above.
\item $|\cos \theta^*_B|$: the cosine of the angle between the $\B$ flight direction and the beam axis in the CM frame.
\item $|\cos \theta^*_T|$: the cosine of the angle between the thrust axis of the $\B$ candidate and the thrust axis of the ROE in the CM frame.
\item $|\cos \theta^*_{T \gamma}|$: the cosine of the angle between the primary photon momentum and the thrust axis of the ROE in the CM frame.
\end{itemize}

To obtain the best sensitivity, we simultaneously optimize, using MC samples, the SSC and BRC selections in four $X_s$ mass ranges ([0.6-1.1], [1.1-2.0], [2.0-2.4], and [2.4-2.8]\gevcc), maximizing $S/{\sqrt{S+B}}$, where $S$ is the number of expected signal events and $\B$ is the number of expected background events with \mes $>$ 5.27 \gevcc.  The optimized selection values are the same for both $\B$ and $\Bb$.

\section{Fitted Asymmetry}
\label{sec:apeak}
For each $\B$ flavor, we describe the \mes distribution with a sum of an ARGUS distribution~\cite{ARGUS}\cite{ARGUS2} and a two-piece normal distribution ($G$)~\cite{BIFGAUSS}:

\begin{align}
	\textrm{PDF}^{b}(\mes) 
	= & \frac{T_{\textrm{cont}}}{2}( 1+A_{\textrm{cont}} )\textrm{ARGUS}(\mes; c^b, \chi^{b}, p^{b} )+ \nonumber \\
	  & \frac{T_{\textrm{peak}}}{2}( 1+A_{\textrm{peak}} )G(\mes; \mu^b, \sigma^b_{L}, \sigma^b_{R}),\\
	\textrm{PDF}^{\bbar}(\mes) 
	= & \frac{T_{\textrm{cont}}}{2}( 1-A_{\textrm{cont}} )\textrm{ARGUS}(\mes; c^{\bbar}, \chi^{\bbar}, p^{\bbar} )+ \nonumber\\
	  & \frac{T_{\textrm{peak}}}{2}( 1-A_{\textrm{peak}} )G(\mes; \mu^{\bbar}, \sigma^{\bbar}_{L}, \sigma^{\bbar}_{R})
\end{align}
where
\begin{align}
	T_{\textrm{cont}} & =  n^b_{\textrm{cont}} + n^{\bbar}_{\textrm{cont}},\\
	T_{\textrm{peak}} & =  n^b_{\textrm{peak}} + n^{\bbar}_{\textrm{peak}}
\end{align}
are the total number of events of both flavors described by the ARGUS distribution and the two-piece normal distribution and
\begin{align}
	A_{\textrm{cont}} & =  \frac{n^b_{\textrm{cont}} - n^{\bbar}_{\textrm{cont}}}{n^b_{\textrm{cont}} + n^{\bbar}_{\textrm{cont}}},\\
	A_{\textrm{peak}} & =  \frac{n^b_{\textrm{peak}} - n^{\bbar}_{\textrm{peak}}}{n^b_{\textrm{peak}} + n^{\bbar}_{\textrm{peak}}}
\end{align}
are the flavor asymmetries of events described by the ARGUS distribution and the two-piece normal distribution, respectively. The superscript $b$ and $\bbar$ indicate whether the parameter belongs to the $b$-quark containing $B$ meson ($\Bzb$ and $\Bm$) distribution or the $\bbar$-quark containing $B$ meson distribution ($\Bz$ and $\Bp$), respectively. In particular, $n^b_{\textrm{peak}}$ and $n^{\bbar}_{\textrm{peak}}$ are the numbers of events in the peaking (Gaussian) part of the distribution. Similarly, $n^b_{\textrm{cont}}$ and $n^{\bbar}_{\textrm{cont}}$ are the numbers of events in the continuum (ARGUS) part of the distribution. The shape parameters for ARGUS distributions are the curvatures ($\chi^b$ and $\chi^{\bbar}$), the powers ($p^b$ and $p^{\bbar}$), and the endpoint energies ($c^b$ and $c^{\bbar}$). The shape parameters for two-piece normal distribution are the peak locations ($\mu^b$ and $\mu^{\bbar}$), the left-side widths ($\sigma_L^b$ and $\sigma_L^{\bbar}$), and the right-side widths ($\sigma_R^b$ and $\sigma_R^{\bbar}$).

\begin{figure*}[htbp]
   \centering
   \includegraphics[width=\textwidth]{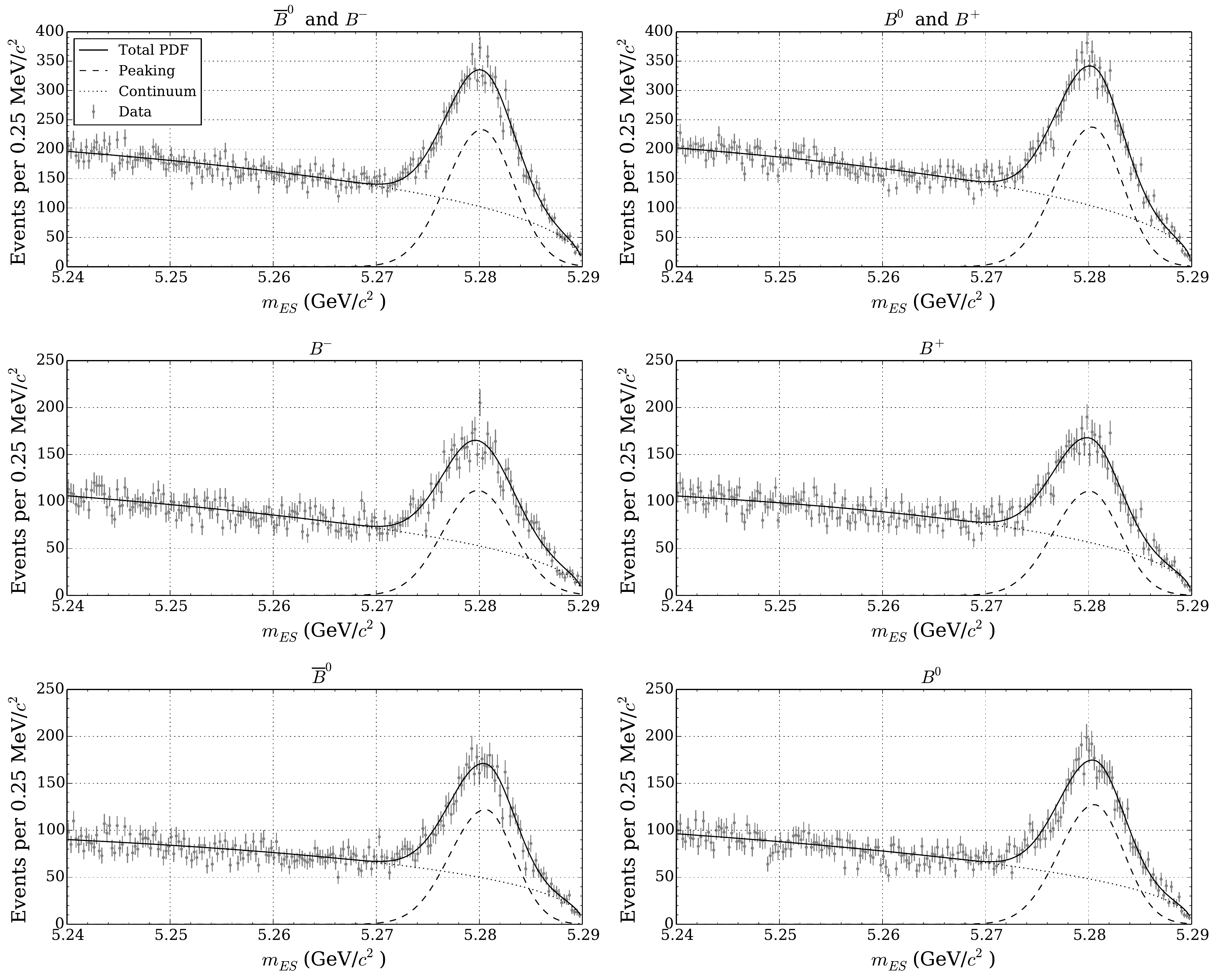} 
   \caption{The \mes distributions along with fitted probability density functions, for: $\Bzb$ and $\Bm$ sample (top left), $\Bz$ and $\Bp$ sample (top right), $\Bm$ sample (middle left), $\Bp$ sample (middle right), $\Bzb$ sample (bottom left), and $\Bz$ sample (bottom right). Data are shown as points with error bars. The ARGUS distribution component, two-piece normal distribution component and the total probability density function are shown with dotted lines, dashed lines, and solid lines, respectively.}
   \label{fig:mes}
\end{figure*}

It should be noted that $A_\textrm{peak}$ is related to \acp defined in Eq.~\ref{eq:acpdef} by the relation shown in Eq.~\ref{eq:acpapeakrelation}. To obtain \apeak, we perform a simultaneous binned likelihood fit for both $\B$ flavors. The ARGUS endpoint energies $c^b$ and $c^{\bbar}$ are fixed at 5.29\gevcc. All other shape parameters for the ARGUS distributions and the two-piece normal distributions are allowed to float separately. Fig.~\ref{fig:mes} shows the \mes distributions, along with fitted shapes. Table~\ref{tab:finalresult} summarizes the results for \apeak.

\section{Detector Asymmetry}
\label{sec:adet}

Part of the difference between \apeak and \acp comes from the difference in $K^+$ and $K^-$ efficiencies. The $K^+$ PID efficiency is slightly higher than the $K^-$ PID efficiency; the difference also varies with the track momentum. The cause of this difference is the fact that the cross-section for $K^-$-hadron interactions is higher than that for $K^+$-hadron interactions. This translates to the $K^-$ having a greater probability of interacting before it reaches the DIRC, thereby lowering the quality of the $K^-$ Cherenkov cone angle measurement, which affects the PID performance.

The first order correction to \acp from $K^+$/$K^-$ efficiency differences is given by
\begin{equation}
	\adet = \frac{\nu_b - \nu_{\bbar}}{\nu_b + \nu_{\bbar}}
	\label{eq:adetdef}
\end{equation}
where $\nu_b$ and $\nu_{\bbar}$ are the number of events for each flavor after all selections, assuming the underlying physics has no flavor asymmetry.

We use a sideband region ($\mes<5.27 \gevcc$) which consists mostly of $e^+e^- \to \q\overline{q}$ events to measure \adet. We do not expect a flavor asymmetry in the underlying physics in this region. We count the number of events in the sideband region for each flavor and use Eq.~\ref{eq:adetdef} to determine $A^\textrm{sideband}_\textrm{det}$.

However, since the difference in $K^-$ and $K^+$ hadron cross section depends on $K$ momentum and the $K$ momentum distributions of the side band region and the peaking region ($\mes > 5.27 \gevcc$) slightly differ, $A^\textrm{sideband}_\textrm{det}$ and \adet need not be identical. The variation of \adet for any $K$ momentum distribution can be bounded by the maximum and minimum value of the ratio between $K^+$ and $K^-$ efficiencies ($\epsilon_{\Kp}/\epsilon_{\Km}$) in the $K$ momentum range of interest:
\begin{equation}
\frac{1}{2}\left(\min_{p_K}\frac{\epsilon_{\Kp}}{\epsilon_{\Km}}-1\right) \leq A_\textrm{det} \leq \frac{1}{2}\left(\max_{p_K} \frac{\epsilon_{\Kp}}{\epsilon_{\Km}}-1\right).
\label{eq:adetbound}
\end{equation}
The final states with no charged $K$ can be considered as having a special value of $p_K$ where $\epsilon_{\Kp}$ and $\epsilon_{\Km}$ are identical.

We use highly pure samples of charged kaons from the decay $D^{*+}\to D^0\pi^+$, followed by $D^0 \to K^-\pi^+$, and its charge conjugate, to measure the ratio of efficiencies for $K^+$ and $K^-$. We find that the deviation from unity of $\epsilon_{\Kp}/\epsilon_{\Km}$ varies from 0 to 2.5\% depending on the track momentum.

The bound given in Eq.~\ref{eq:adetbound} implies that the distribution of the differences between any two detector asymmetries chosen uniformly within the bound is a triangle distribution with the base width of 2.5\%.

The standard deviation of such a distribution is $2.5\%/\sqrt{24} = 0.5\%$. We use $A^\textrm{sideband}_\textrm{det}$ as the central value for $\adet$ and this standard deviation as the systematic uncertainty associated with detector asymmetry. Table~\ref{tab:finalresult} lists the results of $\adet$.

\begin{table*}[htpb]
\caption{Summary of $\acp$ results along with $\adet$ and systematic uncertainties due to peaking background contamination (D) for each \B sample. The \acp's in the last column are calculated using Eq.~\ref{eq:acpapeakrelation}. The first error is statistical, the second (if present) is systematics.}

\setlength{\tabcolsep}{5pt}
\begin{tabular}{c c c c c c }
\hline
\B Sample \TTT\BBB & \apeak & D & \multicolumn{1}{c}{\adet} & \multicolumn{1}{c}{\acp} \\
\hline
\hline
\hline

All \B      & $+(0.33\pm1.87)\%$ & $\pm0.88\%$ & $-(1.40\pm0.49\pm0.51)\%$  & $+(1.73\pm1.93\pm1.02)\%$\\

Charged \B  & $+(3.14\pm2.86)\%$ & $\pm0.80\%$ & $-(1.09\pm0.67\pm0.51)\%$  & $+(4.23\pm2.93\pm0.95)\%$\\

Neutral \B  & $-(2.48\pm2.47)\%$ & $\pm0.97\%$ & $-(1.74\pm0.72\pm0.51)\%$  & $-(0.74\pm2.57\pm1.10)\%$\\

\hline

\hline

\end{tabular}
\label{tab:finalresult}
\setlength{\tabcolsep}{6pt}
\end{table*}%

\section{Peaking Background Contamination}
\label{sec:dilution}

\begin{figure*}[htbp]
   \centering
   \includegraphics[width=\textwidth]{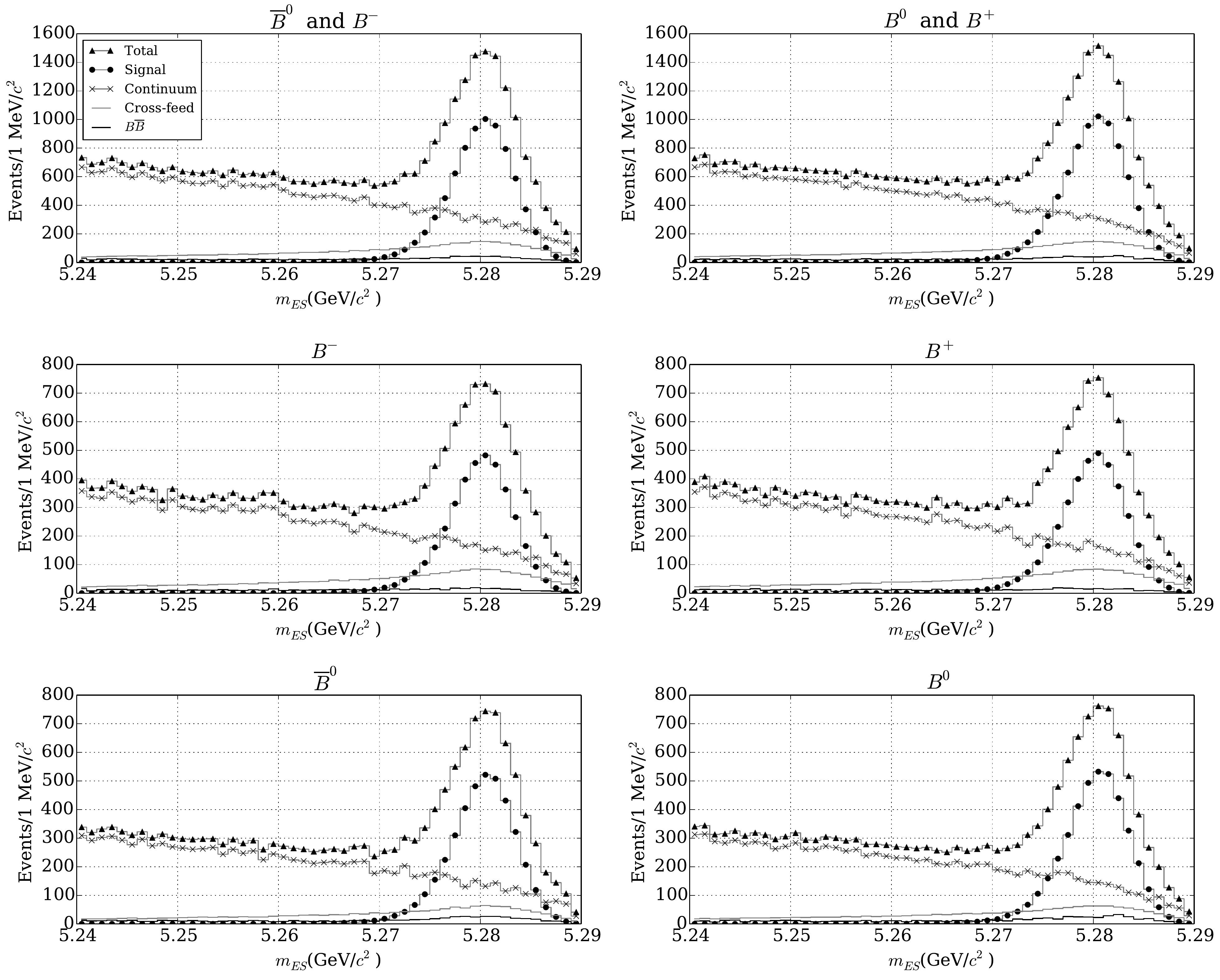} 
   \caption{The contributions to the total \mes distributions (gray lines with triangle markers) from the signal $\BXsg$ (gray lines with circle markers), the continuum background (gray lines with x markers), the cross-feed background (gray lines with no marker), and the generic $\BB$ background (solid black lines) according to the MC sample for: $\Bzb$ and $\Bm$ sample (top left), $\Bz$ and $\Bp$ sample (top right), $\Bm$ sample (middle left), $\Bp$ sample (middle right), $\Bzb$ sample (bottom left), and $\Bz$ sample (bottom right).}
   \label{fig:mcmes}
\end{figure*}

Our fitting procedure does not explicitly separate
the cross-feed and generic $\BB$ backgrounds from the signal.  Both backgrounds
have small peaking components, as shown in Figure~\ref{fig:mcmes}, so the yield for each flavor
used in calculating  $\apeak$ contains both signal and these peaking backgrounds.
We quantify the effect and include it as a source of systematic uncertainty.

Let the number of signal events for $\b$-quark containing $\B$ mesons and $\bbar$-quark containing $\B$ mesons be $n_b$ and $n_{\bbar}$ and the number of contaminating peaking background events misreconstructed as $\b$-quark containing $\B$ mesons and $\bbar$-quark containing $\B$ mesons be $\beta_b$ and $\beta_{\bbar}$. The difference between \apeak and \acp due to peaking background contamination is given by:
\begin{equation}
	D = R \times \delta A \label{eq:ddef},
\end{equation}
where 
$R$ is the ratio of the number of peaking background events to the total number of events in the peaking region, given by
\begin{equation}
R = \frac{\beta_{\b}+\beta_{\bbar}}{n_{\b}+n_{\bbar}+\beta_{\b}+\beta_{\bbar}},
\end{equation}
and $\delta A$ is the difference between the true signal asymmetry and the peaking background asymmetry, given by
\begin{equation}
\delta A = \frac{n_{\b}-n_{\bbar}}{n_{\b}+n_{\bbar}} - \frac{\beta_{b}-\beta_{\bbar}}{\beta_{\b}+\beta_{\bbar}}.
\end{equation}

We estimate $R$ using the MC sample. We use the sum of the expected number of cross-feed background events and expected number of generic $\BB$ events with $\mes > 5.27 \gevcc$ for each flavor as $\beta_b$ and $\beta_{\bbar}$. We obtain $n_b$ and $n_{\bbar}$ from the total number of expected signal events for each flavor. 

Since the peaking background events are from mis-reconstructed \B mesons, the \mes distribution of the peaking background has a very long tail. It resembles the sum of an ARGUS distribution and a small peaking part. The fit to the total \mes distribution is the sum of a two-piece normal distribution and an ARGUS distribution. A significant portion of peaking background is absorbed into the ARGUS distribution causing our estimate of $R$ to be overestimated.

We bound the difference in asymmetry, $\delta  A$, using the range of values predicted by the SM: $-0.6~\%~<~A_{CP}^{SM}~<~2.8~\%$. This gives $|\delta A| < 3.4\%$. This value is also very conservative, since the amount of cross-feed background in the signal region is approximately five times the amount of generic \BB background, and we expect the flavor asymmetry of the cross-feed events to be similar to that of the signal.

We validate our estimates by extracting $\apeak$ from pseudo MC experiments with varying amounts of crossfeed background asymmetry and observe the shift from the true value of the signal asymmetry. The shift is about half the value estimated using the method described; we use the more conservative estimate as our systematic uncertainty. For \acp of the charged and neutral \B, this estimate is conservative enough to cover a large possible range of $|\dacp|<15\%$ that could shift the value of \apeak via the cross-feed of the type $\Bzb\to X_s\gamma$ mis-reconstructed as $\Bm \to \X_s \gamma$ ($\Bzb\Rightarrow\Bm$) and $\Bm\to X_s\gamma$ misreconstructed as $\Bzb \to \X_s \gamma$ ($\Bm\Rightarrow\Bzb$). Table~\ref{tab:contamination} lists the values of $R$, $\delta A$ and $D$.

\begin{table}[htdp]
\setlength{\tabcolsep}{5pt}
\caption{Values of $R$, $\delta A$ and $D$.}
\begin{center}
\begin{tabular}{c c r r r}
 \hline
 \B Sample \TTT\BBB& $R$ & $|\delta A|$ & $D$\\
\hline
\hline
All \B \TTT& 0.26 & 3.4\% & $\pm$0.88\%\\
Charged \B & 0.28 & 3.4\% & $\pm$0.80\%\\
Neutral \B \BBB& 0.24 & 3.4\% & $\pm$0.97\%\\
\hline
\end{tabular}
\end{center}
\label{tab:contamination}
\setlength{\tabcolsep}{6pt}
\end{table}%

\section{Results}
Following Eq.~\ref{eq:acpapeakrelation}, we subtract \adet from \apeak to obtain \acp.  The statistical uncertainties are added in quadrature. Systematic uncertainties from peaking background contamination and from detector asymmetry are added in quadrature to obtain the total systematic uncertainty.  We find
\begin{align}
\acp = +(1.7\pm1.9\pm1.0)\%
\end{align}
where the uncertainties are statistical and systematic, respectively.
Compared to the current world average, the statistical uncertainty is smaller by approximately 1/3 due to the improved rejection of peaking background  described above.

The measurement of \acp is based on the ratio of the number of events, but \acp is defined as the ratio of widths. In order to make the two definitions of \acp equivalent, we make two assumptions. First, we assume that there are as many decaying $\Bz$ mesons as decaying $\Bzb$ mesons, i.e. there is no $\CP$ 
violation in mixing. This has been measured to be at most a few $10^{-3}$ for $B$ mesons~\cite{PDG}. Second, since the ratio of the number of events is essentially the ratio of the branching fractions under the first assumption, we assume that the lifetime of \b- and \bbar-containing \B mesons are identical so that the ratio of the branching fractions is equal to the ratio of the decay widths. This is guaranteed if we assume \CPT invariance. The isospin asymmetry has negligible effect on \acp:  The effect from the difference of $\Bz$ and $\Bp$ lifetime and from \dacp is suppressed by a factor of isospin efficiency asymmetry~\cite{ISOSPINASYMMETRY}, which we find to be on the order of 2\%. The total effect is thus on the order of $10^{-4}$, which is below our sensitivity.

Using the values of \acp for charged \B and neutral \B in Table~\ref{tab:finalresult}, we find
\begin{align}
\dacp =  +(5.0\pm3.9\pm1.5)\%,
\end{align}
where the uncertainties are statistical and systematic, respectively. The statistical and systematic uncertainties on \dacp are obtained by summing in quadrature the uncertainties on the charged and neutral \acp measurements. The systematic uncertainty for \dacp is also validated with an alternative method of estimating the multiplicative effects from the peaking background contamination on \dacp taking into account each component of cross-feed. In particular, the cross-feed of the type $\Bzb \Rightarrow \Bm$, $\Bm \Rightarrow \Bzb$ and generic \BB produce shifts that are proportional to \dacp.  We use a conservative value for the peaking background composition of 2:2:1 ($\Bzb~\Rightarrow~\Bzb~:~\Bm~\Rightarrow~\Bzb$ : Generic \BB) and the value of cross-feed contamination ratio  $R\sim 1/4$. We find the total effect to be conservatively at most $\frac{1}{4}\dacp = 1.3\%$. The estimate is in agreement with the quadrature sum of the peaking background contamination systematics for charged and neutral \B asymmetry, which is $\sqrt{1.0\%^2+0.8\%^2}=1.3\%$.

In the calculation of \dacp, we also assume that the fragmentation does not create an additional asymmetry. This is generally assumed in this type of analysis. This is particularly important for the \dacp measurement since the final states are not all isospin counterparts. With this assumption, we can use  \acp for 10 charged \B final states and \acp for 6 neutral \B final states as \acp for charged \B and neutral \B, respectively.

Using the formula,
\begin{equation}
\dacp \simeq 0.12 \times \frac{\tilde{\Lambda}_{78}}{100\mev} \imes,
\label{eq:dacpim78}
\end{equation}
given in~\cite{GIL}, we can use the measured value of \dacp to determine the 68\% and 90\% confidence limits (CL) on \imes. The interference amplitude, $\tilde{\Lambda}_{78}$, in Eq.~\ref{eq:dacpim78} is only known as a range of possible values,
\begin{equation}
	17\mev < \tilde{\Lambda}_{78} < 190 \mev \label{eq:l78limit}.
\end{equation}

\begin{figure}[htbp]
   \centering
   \includegraphics[width=\columnwidth]{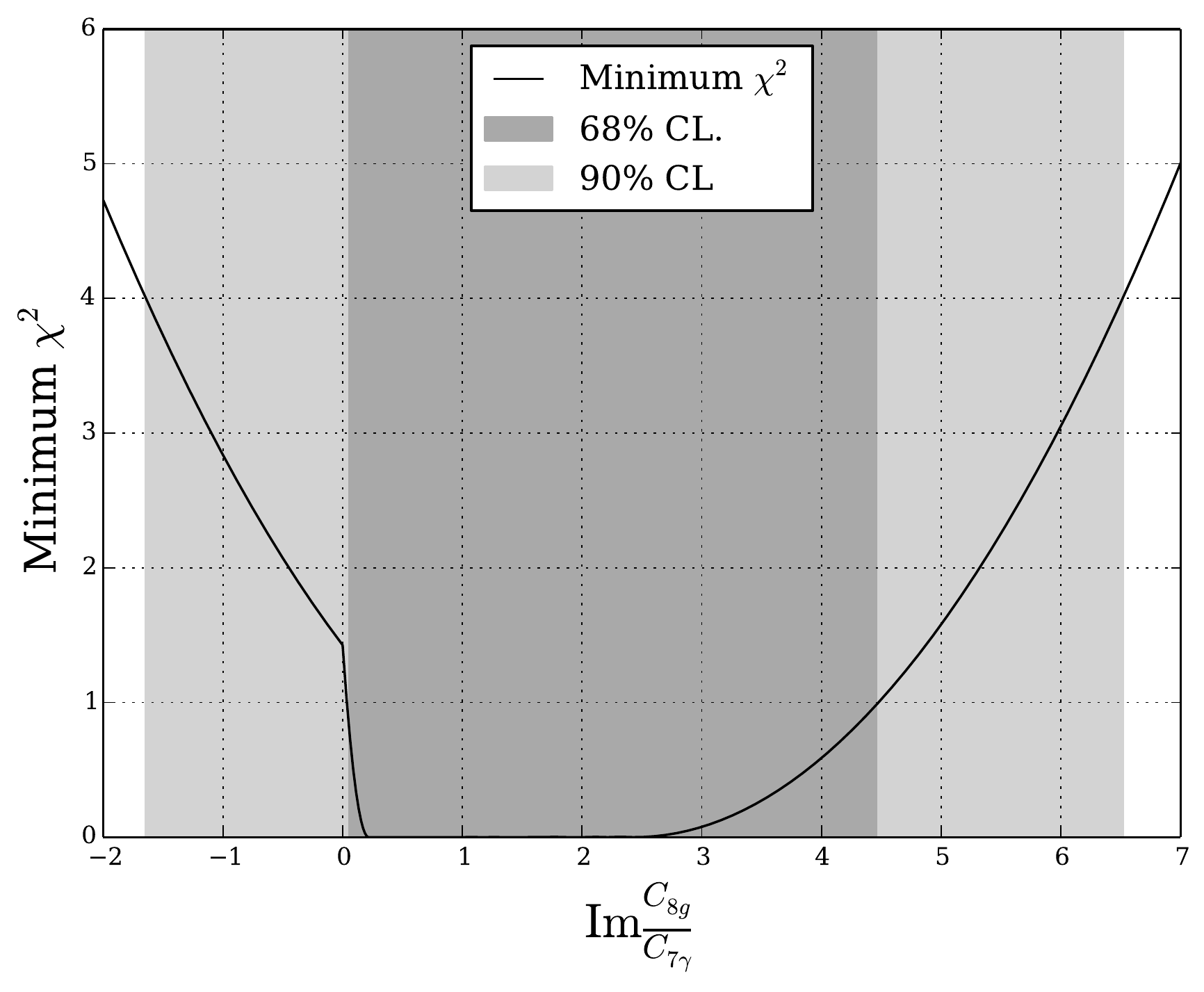} 
   \caption{The minimum $\chi^2$ for given $\operatorname{Im}\frac{C_{8g}}{C_{7\gamma}}$ from all possible values of $\lambdase$. 68\% and 90\% confidence intervals are shown in dark gray and light gray, respectively.}
   \label{fig:im87}
\end{figure}

We calculate a quantity called minimum $\chi^2$ defined by
\begin{equation}
	\textrm{minimum\;} \chi^2 = \frac{\min\limits_{\tilde{\Lambda}_{78}} \left[(\Delta A_{\mathrm{Th}} - \Delta A_{\mathrm{Exp}})^2\right]}{\sigma^2},\end{equation}
where $\Delta A_{\mathrm{Th}}$ is the theoretical prediction of $\dacp$ for given \imes and $\tilde{\Lambda}_{78}$ using Eq.~\ref{eq:dacpim78}, $\Delta A_{\mathrm{Exp}}$ is the measured value, $\sigma$ is uncertainty on the measured value, and the minimum is taken over the range of $\tilde{\Lambda}_{78}$ given in Eq.~\ref{eq:l78limit}. Figure~\ref{fig:im87} shows the plot of minimum $\chi^2$ versus \imes. It has two notable features. First, there is a  plateau of minimum $\chi^2 = 0$. This is the region of \imes where we can always find a value of $\tilde{\Lambda}_{78}$ within the possible range (Eq. \ref{eq:l78limit}) such that $\Delta A_{\mathrm{Th}}$ matches exactly $\Delta A_{\mathrm{Exp}}$. Second, the discontuinity at $\imes = 0$ comes from the fact that the value of $\tilde{\Lambda}_{78}$ that gives the minimum value is different. When \imes is small and positive, we need a large positive $\tilde{\Lambda}_{78}$ to be as close as possible to the measured value, while when \imes is negative, we need a small positive value of $\tilde{\Lambda}_{78}$ to not be too far from the measured value.

\begin{figure}[htbp]
   \centering
   \includegraphics[width=\columnwidth]{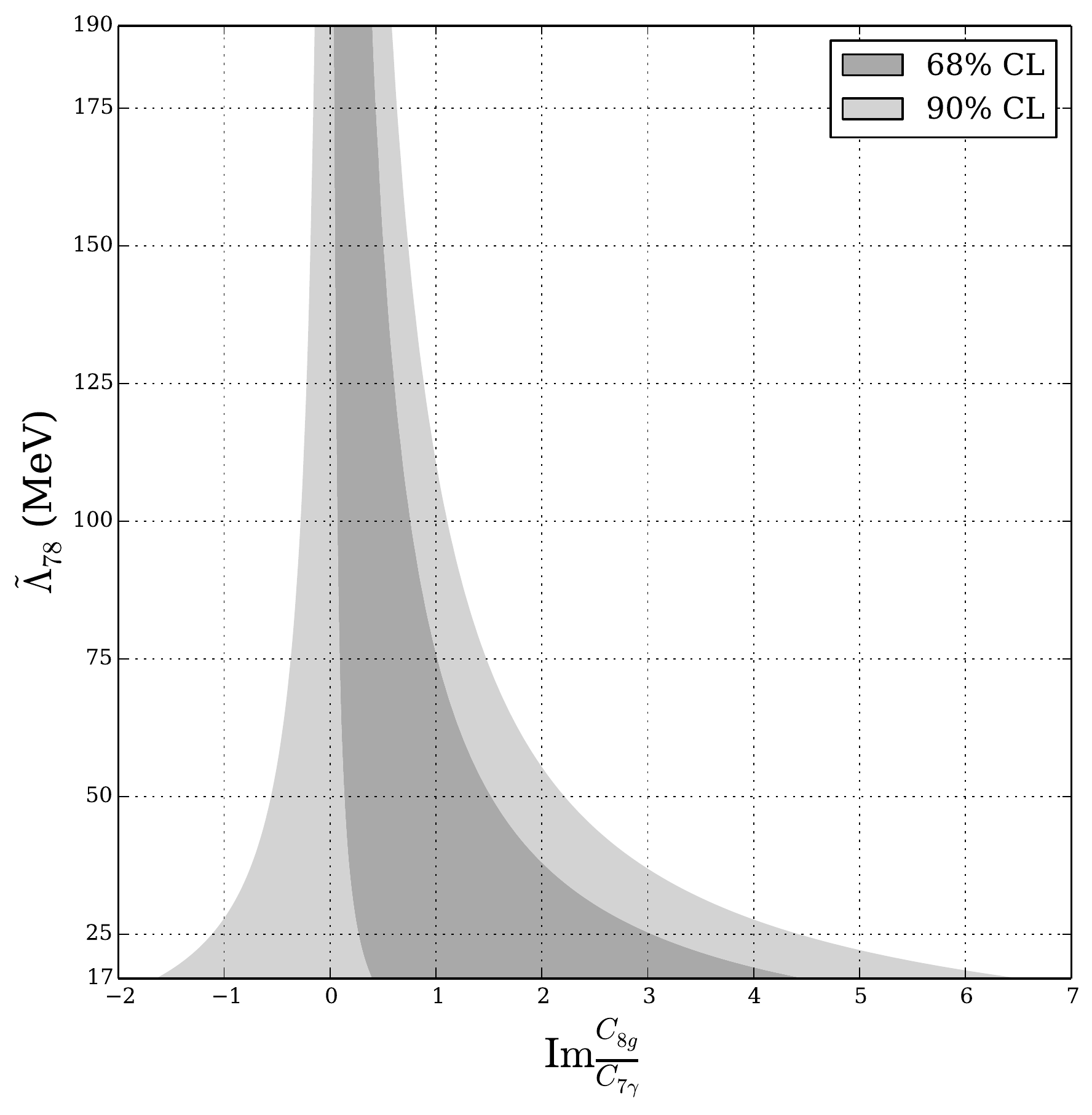} 
   \caption{The 68\% and 90\% confidence intervals for  $\operatorname{Im}\frac{C_{8g}}{C_{7\gamma}}$ and $\lambdase$.}
   \label{fig:im87vl78}
\end{figure}

The 68\% and 90\% confidence limits are then obtained from the ranges of \imes, which yield the minimum $\chi^2$ less than 1 and 4, respectively. We find
\begin{align}
	0.07 &\leq \operatorname{Im} \frac{C_{8g}}{C_{7\gamma}} \leq 4.48, \; \textrm{ 68\% CL},\label{eq:clse}\\
	-1.64 &\leq \operatorname{Im} \frac{C_{8g}}{C_{7\gamma}} \leq 6.52, \; \textrm{ 90\% CL}\label{eq:clnt}.
\end{align}
The dependence of minimum $\chi^2$ on $\imes$ as shown in Figure~\ref{fig:im87} is not parabolic, which would be expected from a Gaussian probability. Care must be taken when combining it with other constraints.
Since the confidence intervals obtained are dominated by the possible values $\lambdase$ at the low end, improvement of limits on $\lambdase$ will narrow the confidence interval. We therefore also provide confidence interval for \imes as the function of $\lambdase$ in Figure \ref{fig:im87vl78}.

\section{Summary}

In conclusion, we present a measurement of the direct \CP violation asymmetry, \acp, in $\B\to X_s\gamma$ and the isospin difference of the asymmetry, \dacp with 429 $\textrm{fb}^{-1}$ of data collected at the $\Upsilon(4S)$ resonance with the \babar\: detector. \B meson candidates are reconstructed from 10 charged $\B$ final states and 6 neutral $\B$ final states. We find $\acp = +(1.7\pm1.9\pm1.0)\%$, in agreement with the SM prediction and with the uncertainty smaller than that of the current world average. We also report the first measurement of $\dacp =  +(5.0\pm3.9\pm1.5)\%$, consistent with the SM prediction. Using the value of \dacp, we calculate the 68\% and 90\% confidence intervals for $\imes$ shown in Eqs.~\ref{eq:clse} and Eq.~\ref{eq:clnt}, respectively. The confidence interval can be combined with existing constraints on $C_{7\gamma}$ to provide a constraint on $C_{8g}$.

\section{acknowledgements}
We would like to thank Gil Paz for very useful discussions.
We are grateful for the 
extraordinary contributions of our \pep2\ colleagues in
achieving the excellent luminosity and machine conditions
that have made this work possible.
The success of this project also relies critically on the 
expertise and dedication of the computing organizations that 
support \babar.
The collaborating institutions wish to thank 
SLAC for its support and the kind hospitality extended to them. 
This work is supported by the
US Department of Energy
and National Science Foundation, the
Natural Sciences and Engineering Research Council (Canada),
the Commissariat \`a l'Energie Atomique and
Institut National de Physique Nucl\'eaire et de Physique des Particules
(France), the
Bundesministerium f\"ur Bildung und Forschung and
Deutsche Forschungsgemeinschaft
(Germany), the
Istituto Nazionale di Fisica Nucleare (Italy),
the Foundation for Fundamental Research on Matter (The Netherlands),
the Research Council of Norway, the
Ministry of Education and Science of the Russian Federation, 
Ministerio de Econom\'{\i}a y Competitividad (Spain), the
Science and Technology Facilities Council (United Kingdom),
and the Binational Science Foundation (U.S.-Israel).
Individuals have received support from 
the Marie-Curie IEF program (European Union) and the A. P. Sloan Foundation (USA). 



\begin{thebibliography}{99}
\bibitem{KLFraction}
If we include $K_L$ modes as if they have same branching fraction as $K_S$ modes, the final state coverage for $0.6 < m_{\xs} < 2.0 \gevcc$ is 69\% for charged \B and 34\% for neutral \B.
\bibitem{lateralmoment}
The lateral moment is the ratio for the sum of energies of all but the two most energetic crystals in the cluster weighted by the squares of distances to the cluster center and the sum of energies of all crystals weighted by the square of distance to the cluster center.
\bibitem{Thrust}
\[
	\textrm{Thrust} = \max_A \left[ \frac{\sum\limits_{i=1}^N \left| A \cdot p_i \right|}{\sum\limits_{i=1}^N \sqrt{ p_i \cdot p_i}} \right]
\]
where $A$ is a unit vector, $p_i$ are three momenta of the decay particles of the $B$ candidate.
\bibitem{BIFGAUSS}
\begin{align*}
\textrm{G}(x; \mu, \sigma_L, \sigma_R) =& N \times
	\begin{cases}
	\exp \left\{ \frac{(x-\mu)^2}{2\sigma_L^2}\right\} & \text{if } x<\mu\\
	\exp \left\{ \frac{(x-\mu)^2}{2\sigma_R^2}\right\} & \text{if } x\geq\mu\\
	\end{cases}
\end{align*}
where $N$ is the normalization.
\end{thebibliography}
\end{document}